\documentclass{article}

\PassOptionsToPackage{numbers, compress}{natbib}

\usepackage[preprint]{neurips_data_2023}

\usepackage[utf8]{inputenc} 
\usepackage[T1]{fontenc}    
\usepackage{hyperref}       
\usepackage{url}            
\usepackage{booktabs}       
\usepackage{amsfonts}       
\usepackage{nicefrac}       
\usepackage{microtype}      
\usepackage{xcolor}     

\usepackage{amsthm,amsmath,amssymb}
\usepackage{mathrsfs}
\usepackage{graphicx}
\usepackage{subfigure}

\usepackage{wrapfig}
\usepackage{stfloats}
\usepackage{bbding}

\usepackage{setspace} 
\usepackage{pifont} 
    
\usepackage{fontawesome}  %
\usepackage{epsfig}
\usepackage{multirow}
\usepackage{booktabs}
\usepackage{svg}
\usepackage{pifont}
\usepackage{amssymb}
\usepackage{enumitem}

\newcommand{\xmark}{\ding{55}}

\title{Towards Long Form Audio-visual Video Understanding}

\author{
Wenxuan Hou\textsuperscript{1,$\dagger$}, 
Guangyao Li\textsuperscript{1,$\dagger$}, 
Yapeng Tian\textsuperscript{2}, 
Di Hu\textsuperscript{1,}\thanks{
Corresponding author.
\textsuperscript{$\dagger$}Equal contribution. 
}
\vspace{0.5em}
\\
\textsuperscript{1}Gaoling School of Artificial Intelligence, Renmin University of China\\
\textsuperscript{2} Department of Computer Science, The University of Texas at Dallas
\\
\textsuperscript{1}\{wxhou, guangyaoli, dihu\}@ruc.edu.cn, \textsuperscript{2}\{yapeng.tian\}@utdallas.edu
\vspace{-1.5em}
}

\begin{document}

\maketitle

\begin{abstract}
We live in a world filled with never-ending streams of multimodal information. As a more natural recording of the real scenario, long form audio-visual videos are expected as an important bridge for better exploring and understanding the world.
In this paper, we propose the multisensory temporal event localization task in long form videos and strive to tackle the associated challenges. 
To facilitate this study, we first collect a large-scale Long Form Audio-visual Video (LFAV) dataset with \textcolor{black}{5,175 videos and an average video length of 210 seconds. Each of the collected videos is elaborately annotated with diversified modality-aware events, in a long-range temporal sequence}. We then propose an event-centric framework for localizing multisensory events as well as understanding their relations in long form videos. It includes three phases in different levels: \textcolor{black}{snippet prediction phase to learn snippet features, event extraction phase to extract event-level features, and event interaction phase to study event relations}.
Experiments demonstrate that the proposed method, utilizing the new LFAV dataset, exhibits considerable effectiveness in localizing multiple modality-aware events within long form videos.
\textcolor{black}{Project website: \href{http://gewu-lab.github.io/LFAV/}{http://gewu-1ab.github.io/LFAV/}}
\end{abstract}

\section{Introduction}

Guiding the machine to perceive and understand natural scenes like human beings is a long term vision of the AI community. As a primary means for recording natural scenes, video plays an important role in approaching the above prospect. Previous works on video understanding have achieved considerable performance in many tasks, including action recognition~\cite{simonyan2014two,wang2016temporal,feichtenhofer2019slowfast,lin2019tsm}, temporal action localization (TAL)~\cite{shou2016temporal,yuan2016temporal,chao2018rethinking,zeng2019graph}, weakly supervised temporal action localization (WTAL)~\cite{nguyen2018weakly,shou2018autoloc,yang2022acgnet,huang2022weakly}, audio-visual event localization (AVE)~\cite{tian2018audio,wu2019dual,zhou2021positive}, audio-visual video parsing (AVVP)~\cite{tian2020unified,wu2021exploring,jiang2022dhhn}, \emph{etc}. However, videos they studied are usually pre-processed (\emph{e.g.} trimming long videos into short meaningful clips or ignoring other modalities such as audio), which could result in the
potential deviation when depicting the real scenes.

\begin{figure*}[t]
\centering
\includegraphics[width=0.95\linewidth]{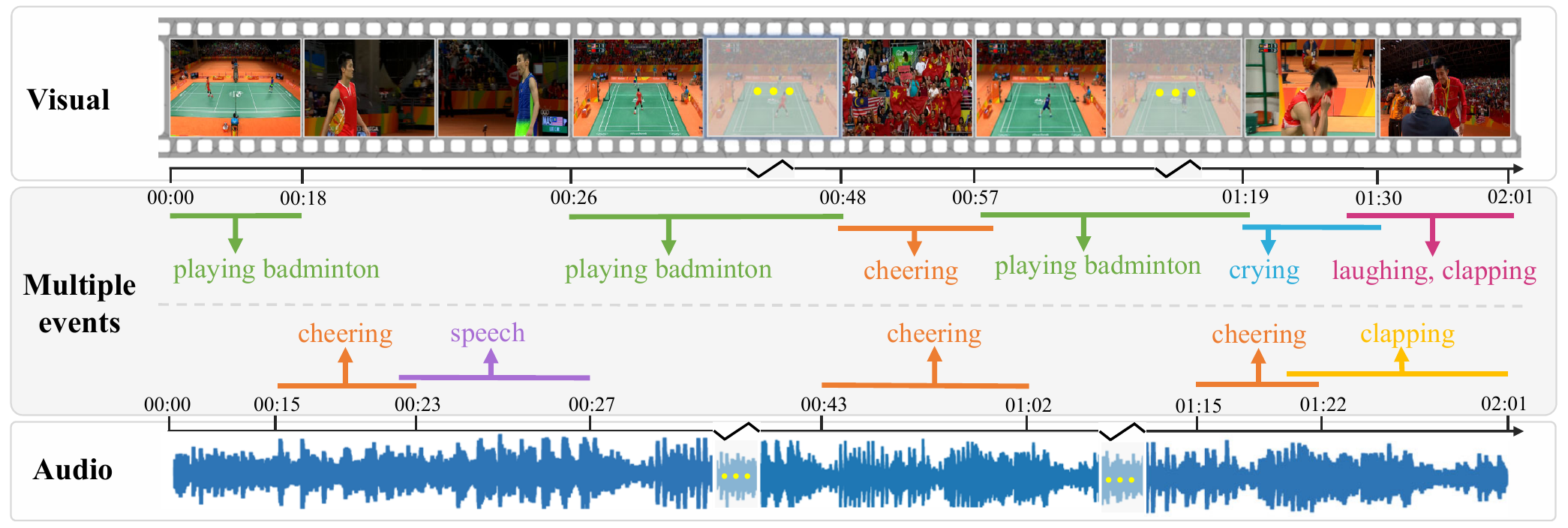}
\vspace{-0.5em}
\caption{A long form audio-visual video example, with a length of 121 seconds. This video shows a badminton game. The audio modality contains three events: \emph{cheering}, \emph{clapping} and \emph{speech}, the visual modality contains five events: \emph{playing badminton}, \emph{cheering}, \emph{crying}, \emph{laughing}, and \emph{clapping}. The event \emph{cheering} and \emph{clapping} appears both in audio modality and visual modality. These multisensory events jointly facilitate the understanding of this video.
}
\vspace{0.5em}
\label{fig:teaser}

\end{figure*}

Videos captured from natural scenes have two typical characteristics: 
1) Long form. They usually span several minutes, covering multiple related events in different categories. These events usually jointly contribute to depicting the main content of the video.
2) Audio-visual. 
Videos recorded in real-world scenarios usually comprise both audio and visual modalities. These two aspects often exhibit asynchrony, providing unique perspectives in delineating the video content, yet collaboratively facilitating video understanding.
Figure~\ref{fig:teaser} illustrates a video example of \textcolor{black}{\emph{a badminton game}}. \textcolor{black}{This video is 121-second long, consisting of various audio events and visual events, they 
either only occur in one modality or occur in both modalities but have different temporal boundaries.} These modality-aware events, as well as their inherent relations, help to effectively infer what happens in the video and then achieve a better understanding of the video content. 
Considering the merits of the above two characteristics, we propose to study video understanding in terms of long form and audio-visual aspect, name as \emph{long form audio-visual video understanding}. 

To achieve a better understanding of long form audio-visual videos, we propose to focus on the \emph{multisensory temporal event localization} task, which essentially requires the model to predict the start and end time of each audio and visual event \textcolor{black}{in the video}. However, there remain several challenges when addressing this task.
\textcolor{black}{Firstly, the video contains multiple events with diverse categories, modalities, and varying lengths. 
Secondly, understanding the video content requires effectively modeling long-range dependencies and relations across different clips and modalities.}
To study the above new task, we elaborately build a large-scale Long Form Audio-visual Video (LFAV) dataset with 5,175 videos, as existing datasets are not appropriate for our proposed task. \textcolor{black}{Specifically, existing datasets either just localize audio-visual events (\emph{i.e.,} events that are both audible and visiable~\cite{tian2018audio}) or only localize events in short trimmed videos.} \textcolor{black}{We annotate 24,875 modality-aware event labels on video-level in total.} For validation and testing sets, we annotate the category and temporal boundaries of 23,666 audio and visual events. 
The total length, average length, and average event categories of the videos in LFAV are 302-hour, 210-second, and 3.15, respectively. We expect the LFAV dataset could facilitate the study of our proposed task in the long form audio-visual context.

To address the above challenges, we propose an event-centric framework containing three phases from snippet prediction\footnote{As usual, a snippet represents a 1-second long video segment\cite{tian2020unified}.}, event extraction to event interaction. 
\textcolor{black}{Firstly, we propose a pyramid multimodal transformer model to learn snippet-level features by executing intra-modal and cross-modal interaction within multiscale temporal windows.
Secondly, we extract event-level features by refining and aggregating event-aware snippet features in structured graphs.
At last, we study event relations by modeling the influence among multiple audio and visual events, then refining the event features.}
The three phases are jointly optimized with video-level event labels in an end-to-end fashion. Extensive experimental results show that our event-centric framework can achieve effective multisensory temporal event localization, \textcolor{black}{enhancing understanding of long form audio-visual videos and advancing realistic scene perception.} To summarize, our contributions are threefold:
\vspace{-0.35em}
\begin{itemize}[leftmargin=*]
\item We direct that long form and audio-visual are two key characteristics of videos from natural scenes, then propose to explore long form audio-visual video understanding by concentrating on the proposed multisensory temporal event localization task.
\vspace{-0.35em}
\item We collect a large-scale long form audio-visual video dataset, named LFAV to facilitate our study, which contains 5,175 videos with an average length of above 210 seconds, average event categories of 3.15 per video, and modality-aware annotations.
\vspace{-0.35em}
\item We propose an event-centric framework to tackle multisensory temporal event localization. Experimental results show that our framework obviously surpasses comparison methods, which indicates the effectiveness of the proposed event-centric framework.
\end{itemize}

\begin{figure*}[!t]
 \centering
 \includegraphics[width=0.88\textwidth]{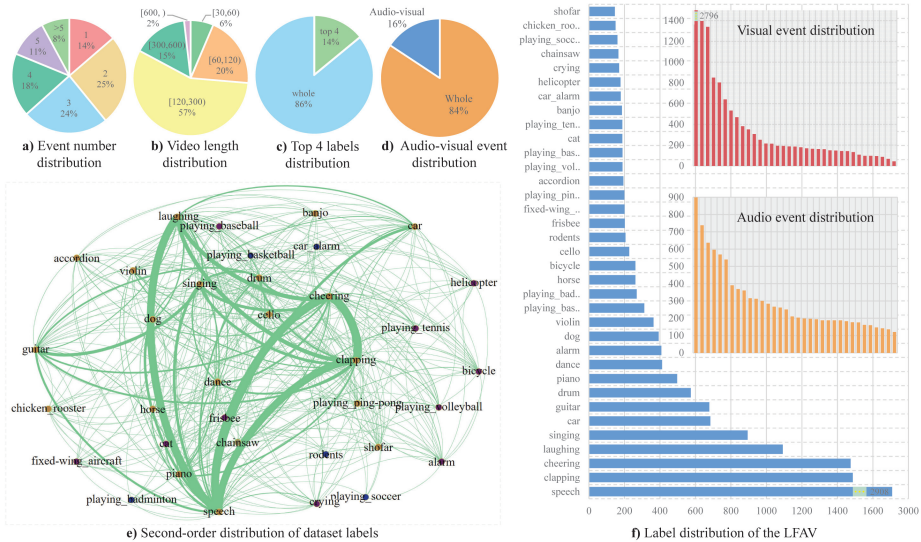}
 \vspace{-0.5em}
 \caption{\textbf{Illustrations of our LFAV dataset statistics}. \textbf{(a-d)} Statistical analysis of label categories, \textcolor{black}{including the distribution of event numbers in each video; the distribution of video length; the proportion of the top 4 event categories, the \textit{top 4 labels} represent \textit{speech, clapping, cheering}, and \textit{laughing}, which are the most common human actions; the temporal proportion of events that occur on two modalities at the same time.} \textbf{(e)} Second-order interactions between all labels, the thicker the line, the closer the association. 
 \textbf{(f)} Distribution of dataset labels \textcolor{black}{of each category}.}
 \label{fig:labels_dist}
\vspace{-1em}
\end{figure*}

\section{Related Work}
\label{gen_inst}
\textbf{Long Form Video Understanding.}
The human vision system is one of the most important bridge for us to perceiving the world~\cite{marr2010vision}. Efforts have been made to understand visual content effectively~\cite{deng2009imagenet,he2016deep,carreira2017quo,liu2022video} and efficiently~\cite{sandler2018mobilenetv2,lin2019tsm}. However, they mainly concentrate on images and short video clips, which struggle to accurately reflect the full picture of the ever-changing real world. In contrast, long form videos provide richer information, enabling the learning of event relations and long-range temporal dependencies.
In recent years, several benchmarks of long form videos have been proposed~\cite{wu2021towards,caba2015activitynet,tang2019coin,zhao2019long,soldan2022mad,zhang2023logo}. Their task definitions vary, but all require a comprehensive understanding of the entire long-form video. Many existing works that also have explored understanding long form videos in multiple views, including improving model architectures~\cite{shou2016temporal,wang2017untrimmednets,shou2017cdc,xu2021long,wu2022memvit,zhang2022actionformer}, learning key clues or regions in the video~\cite{fish2022two,wu2021towards}, aligning cross-modal information~\cite{sun2022long}, \emph{etc.} 
However, previous long-form video benchmarks and learning methods have largely ignored the audio signal, which can provide valuable or visually invisible information. Ignoring the audio channel can lead to an incomplete or biased understanding of long-form videos.

\vspace{-0.15em}
\textbf{Audio-Visual Video Understanding.}
Inspired by the multisensory perception of humans, the community has paid more and more attention to audio-visual scene understanding in recent years. Existing works mainly contain audio-visual action recognition~\cite{kazakos2019epic, xiao2020audiovisual,gao2020listen,chen2022mm}, audio-visual question answering\cite{alamri2019audio,yun2021pano,li2022learning}, \
\textcolor{black}{egocentric audio-visual object localization~\cite{huang2023egocentric}, }
audio-visual segmentation~\cite{zhou2022audio}, \emph{etc}. Recently, the fine-grained temporal event localization task was introduced in the audio-visual learning community, including AVE~\cite{tian2018audio,wu2019dual,zhou2021positive,bagchi2021hear}, AVVP~\cite{tian2020unified,wu2021exploring,jiang2022dhhn}, \textcolor{black}{and dense-localizing audio-visual events~\cite{geng2023dense}. These works aim to temporally recognize events in audio-visual videos.} 
\textcolor{black}{However, the AVE task and the dense-localizing audio-visual events task focus exclusively on events that are both audible and visible, overlook the prevalence of modality-aware events in real-world scenes. The AVVP
task takes modality-aware events into account, but just
localizes events in 10-second short trimmed clips. Besides these audio-visual event localization works, some other works also both take audio and visual modalities into account~\cite{bagchi2021hear,soldan2022mad,EPICSOUNDS2023}, but they mainly aim to boost the uni-modal learning under the assistance of another modality. These works are somewhat lack of the exploration of complex relations among multiple modality-aware events as well as long-range dependencies, hindering a complete and realistic understanding of natural scenarios.}
Compared with these previous works, our multisensory temporal event localization task is more challenging and closer to the real scene, in terms of the properties of long-range and modality-aware events.

\begin{table*}[]
\begin{center}

\caption{\textbf{Comparison with other datasets.} Our LFAV dataset is collected for the proposed multisensory temporal event localization task, where diversified domains are covered. Specifically, the LFAV dataset offers modality-aware annotations for each video, that is it points out the events are from audio, visual, or both modalities. Meanwhile, multiple events with different semantic categories per video are also annotated for better exploring the relation among events. Videos in the dataset have an average length of 210 seconds and a total length of 302 hours. $^\ast$: LLP only provides modality-aware annotations in \textcolor{black}{validation and testing sets}.}
\scalebox{0.72}{
\begin{tabular}{ccccccccc}
\toprule
      &    &    &     &    \\ 
\multirow{-2}{*}{\textbf{Dataset}}  & 
\multirow{-2}{*}{\textbf{Domain}}   & 
\multirow{-2}{*}{\textbf{Year}}  & 
\multirow{-2}{*}{\textbf{\begin{tabular}[c]{@{}c@{}}Modality aware\\annotations \end{tabular}}}   & 
\multirow{-2}{*}{\textbf{\begin{tabular}[c]{@{}c@{}}\ Multiple event\\ annotations \end{tabular}}} &

\multirow{-2}{*}{\textbf{\begin{tabular}[c]{@{}c@{}}Avg\\ Length (\textit{sec.})\end{tabular}}} &
\multirow{-2}{*}{\textbf{\begin{tabular}[c]{@{}c@{}}Total\\ hours\end{tabular}}} \\ \midrule

GTEA~\cite{fathi2011learning} & Human & CVPR'11  & \xmark & \checkmark  & 130    & 0.58      \\
50 Salads~\cite{stein2013combining}& Human & UbiComp'13    & \xmark & \checkmark & 324  & 4.5     \\
Breakfast~\cite{kuehne2014language} & Kitchens & CVPR'14   & \xmark & \checkmark & 162   & 77         \\
ActivityNet~\cite{caba2015activitynet} & Sports & CVPR'15 & \xmark  & \xmark  & 113     & 648          \\
Charades~\cite{sigurdsson2016hollywood}  & Indoor activities &  ECCV'16   & \xmark & \checkmark  & 30       & 82  
\\
EPIC-KITCHENS-100~\cite{damen2022rescaling}& Kitchens  & IJCV'22  & \xmark & \checkmark & 514    & 100    \\

\midrule
Long-Form VQA~\cite{zhao2019long}  & Sports  &  TIP'19 &\xmark  & \xmark  & 128  & 849    \\

LVU~\cite{wu2021towards}  & Movies & CVPR'21 & \xmark  & \xmark & $<$ 180  & N/A  \\
\textcolor{black}{LOGO}~\cite{zhang2023logo}  & Sports & CVPR'23 & \xmark  & \checkmark & 204.2  & 11.3  \\

\midrule

AVE~\cite{tian2018audio}     & Sound event & ECCV'18  & \xmark & \xmark  & 10     & 11        \\
VGGSound~\cite{chen2020vggsound} & In the wild & ICASSP'20 & \xmark  & \xmark &   10  & 560 \\
LLP~\cite{tian2020unified}& Sound event  
 & ECCV'20   & \xmark$^{\ast}$ & \checkmark & 10   & 32.9  \\

\textcolor{black}{UnAV-100}~\cite{geng2023dense}& In the wild  
 & CVPR'23   & \xmark & \checkmark &  42.1  & 126.2  \\

\hline
{{LFAV (Ours)}} & \begin{tabular}[c]{@{}c@{}}Sports,human,\\ instruments, etc.\end{tabular} &
{\textbf{-}} &
{\textbf{\checkmark}} &{\textbf{\checkmark}}& {{210}}  & {{302}}    \\ \bottomrule
\end{tabular}
}
\vspace{-1em}

\label{table:comp_data}
\end{center}
\vspace{-0.5em}
\end{table*}

\section{The LFAV Dataset}
\label{others}

\subsection{Overview}
\label{ssec:3.1}

Towards long form audio-visual video understanding, we build a large-scale audio-visual video dataset, named LFAV. As noted above, high-quality datasets are of considerable value for audio-visual video understanding research. The built LFAV contains 5,175 untrimmed YouTube videos spanning over 35 categories. A wide range of events (\emph{e.g.,} \emph{sing}, \emph{crying}, \emph{playing badminton} and \emph{chainsaw} \emph{etc}.) from diverse domains (\emph{e.g.,} human activities, tools, instrument, sports, and traffic \emph{etc}.) are included.

As shown in Tab.~\ref{table:comp_data}, compared to existing related datasets, our built LFAV dataset has the following advantages: 
\textbf{1)} Videos in the LFAV dataset are usually several minutes long, the value of long-range dependencies and relations among multiple events with different lengths can be explored adequately.
In contrast, \textcolor{black}{most} previous audio-visual datasets~\cite{tian2018audio,tian2020unified,chen2020vggsound} are built with trimmed videos with only 10-second long, which are limited in exploring and utilizing the above properties of long form videos.
\textbf{2)} The LFAV dataset contains videos with multisensory events and modality-aware annotations, which provides the possibility to explore the influence and associations between audio and visual events. Previous video understanding datasets~\cite{fathi2011learning,damen2022rescaling,caba2015activitynet, wu2021towards} mainly focus on visual content perception. Although some of them, such as ActivityNet~\cite{caba2015activitynet},
EPIC-KITCHENS-100~\cite{damen2022rescaling}, \textcolor{black}{as well as some audio-visual datasets~\cite{tian2018audio,geng2023dense},} also offer audio recordings, they do not provide audio-level annotations.
More seriously, the audio information is \textcolor{black}{sometimes} accompanied by severe noise (\textit{e.g.} background music).
In contrast, our LFAV dataset avoids the above issues and thus better supports the study on audio-visual video understanding.

\subsection{Video Collection}

\label{ssec:3.2}
We collect 5,175 videos from YouTube, covering five kinds of daily life to ensure the diversity, complexity, and dynamic of the real world: human-related, sports, musical instruments, tools, and animals. We also construct a label set of 35 kinds of events covering the above scenes, \textcolor{black}{as shown in Tab.~\ref{table:lcc}. 
To keep the balance of different labels, we design the following three collecting steps: }

\textcolor{black}{
\textbf{1) Video retrieval}.
The rule of this step is that each label is required to combine with one or more labels in different categories.
Specifically, we use permutation and combination methods for 35 labels to ensure that all label combinations can be covered in the video as much as possible.}

\textcolor{black}{
\textbf{2) Video filtering}.
Video collectors watch the entire video under the condition of sound playback to select high-quality videos. Low-quality videos, such as videos with audio noise, or excessively trimmed videos, will not be selected to construct the dataset.
In addition, collectors also need to record the categories appearing in both audio and visual modalities.}

\textcolor{black}{
\textbf{3) Label distribution regulation}.
To avoid seriously long-tailed distributions, we need to control the number of each event category. We count the interaction between different event categories when we have collected a certain amount of videos. Then the number of each event category and the confusion matrix among different categories will be used to guide further data collection.}

\begin{table}[]
\begin{center}
\vspace{-0.5em}
\caption{List of 35 label categories, which belong to 5 different kinds of daily life scenes.}
\scalebox{0.95}{

\begin{tabular}{c|c}
\hline
\textbf{Daily life scene}  & 
\textbf{Label categories}                                                                \\ \hline

Musical instruments & guitar, drum, violin, piano, accordion, banjo, cello, shofar \\

\hline
Human-related & speech, singing, crying, laughing, clapping, cheering, dance \\

\hline
Animals & dog, cat, chicken rooster, horse, rodents \\
\hline
Traffic and tools & car, helicopter, fixed-wing aircraft, bicycle, alarm, chainsaw, car alarm \\

\hline
Sports   & \begin{tabular}[c]{@{}c@{}}playing basketball, playing badminton,  
            playing volleyball, \\ playing tennis, 
            playing ping-pong, frisbee, playing soccer\end{tabular} \\ 

\hline
\end{tabular}

}

\label{table:lcc}
\end{center}
\vspace{-1.5em}
\end{table}

\subsection{Annotations}

\label{ssec:3.3}
\textcolor{black}{
For the collected daily life videos, the event annotations contain two parts: \textbf{video-level} annotations (annotate audio event categories and visual event categories that exist in the video) and \textbf{event-level} annotations (annotate temporal boundaries of events in the video.). For most videos in the LFAV dataset (3,721 videos), we just annotate them at video-level, these videos are used as the training set. For a small part of the videos (1,454 videos), we both annotate them at video-level and event-level, these videos are used as the validation set and testing set.}

\textcolor{black}{
To ensure the quality of video annotations and improve the reliability of evaluation results, we verify all annotations manually from two aspects: \textbf{1)} All annotations are randomly assigned to annotators for one-to-one proofreading to check whether existing misplaced annotations, missing annotations, and videos with excessive noise, \textit{etc}.
\textbf{2)} Spot-check the verified annotation information via sampling inspection. After verification, videos with wrong annotations will be re-annotated by other annotators, and videos with excessive noise will be directly deleted.}

\subsection{Statistical Analysis}

\label{ssec:3.4}

Our LFAV dataset contains 5,175 videos spanning over 35 categories for over 302 hours. Fig.~\ref{fig:labels_dist}(a-d) provides the statistical analysis of our dataset. In this dataset, more than half of the videos contain at least three event categories, \textcolor{black}{indicating that there are widely diverse categories of events present in long form audio-visual videos}. All the videos are longer than 30-second and each of them has audio or visual events of at least 1 second. 74\% of the videos are longer than two minutes, \textcolor{black}{and 2\% of the videos are even longer than ten minutes.} 
Fig.~\ref{fig:labels_dist}(e) shows the interaction situation among all the labels, where the thicker the line, the more the interaction.
Fig.~\ref{fig:labels_dist}(f) shows the number of label categories, and the occurrences of each category are no less than 146.
Before annotations, we randomly split the dataset into training, validation, and testing sets with 3,721, 486, and 968 videos, respectively.
Finally, we have 24,875 video-level event annotations \textcolor{black}{on the whole dataset}, and 23,666 second-level event annotations on validation and testing sets in total. 
More details about the statistical analysis of the LFAV dataset are in \textcolor{black}{Sec. B} of the \textit{Supp. Materials.}

\begin{figure*}[!t]
\centering
\includegraphics[scale=0.69]{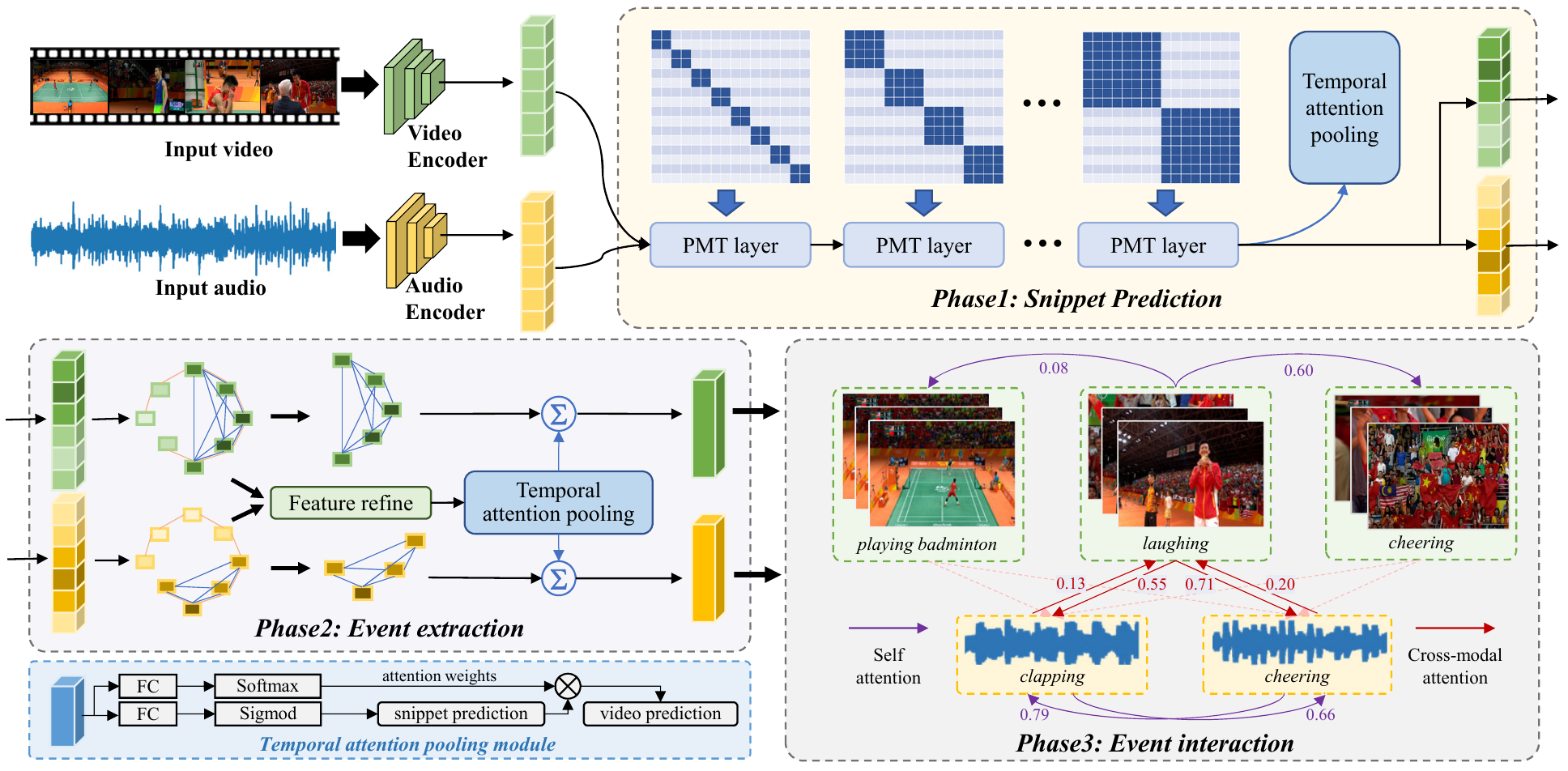}
\vspace{-1em}
\caption{\textbf{Our event-centric framework.} \textbf{Top:} In the first phase of snippet prediction, we propose a pyramid multimodal transformer to generate the snippet features as well as their category prediction. \textbf{Middle left:} In the second phase of event extraction, we build an event-aware graph to refine the snippet features and then aggregate the event-aware snippet features into event features. \textbf{Bottom right:} In the third phase of event interaction, we model the event relations in both intra-modal and cross-modal scenarios and then refine the event feature by referring to its relation to other events. \textbf{Bottom left:} The architecture of temporal attention pooling.}
\label{fig:framework}
\vspace{-1.5em}
\end{figure*}

\section{The Multisensory Temporal Event Localization Task}
\textbf{Task Definition.} The multisensory temporal event localization task in long form audio-visual videos aims at precisely localizing modality-aware events in videos with several minutes long. 
\textcolor{black}{Some} previous works on weakly supervised temporal action localization divide the video into several non-overlapping snippets, then snippets are classified by multiple instance learning and aggregated to events~\cite{wang2017untrimmednets, yang2022acgnet}. We follow this paradigm to define the task, the output of the task is event categories of all snippets. Then event-level predictions can be directly generated by concatenating consecutive snippets with the same class of events. Concretely, for a long form video with a length of $T$-second, we first divide it into $T$ non-overlapping audio snippets $\{{A_t}\}^T_{t=1}$ and visual snippets $\{{V_t}\}^T_{t=1}$, each audio and visual snippet is 1-second long and $T$ is usually up to several hundred. The goal of the task is to generate snippet-level event predictions $\{{p^a_t}\}^T_{t=1}$ and $\{{p^v_t}\}^T_{t=1}$, where $p^a_t\in R^C$ and $p^v_t\in R^C$, they are audio and visual predictions of the $t$-th snippets, respectively. $C$ is the number of categories. The audio event labels ${y^a} \in \left\{0,1\right\} ^C$ and visual event labels ${y^v} \in \left\{0,1\right\} ^C $ are both available during training, they are multi-label binary vectors that indicate the categories of contained modality-aware events in video-level but without snippet-level timestamps.

\textbf{\textcolor{black}{Evaluation Metrics.}} To achieve a more comprehensive evaluation, we use two metrics to evaluate snippet-level and event-level performance, respectively. For snippet-level, we use mAP as the evaluation metric because snippet-level prediction can be regarded as a multi-label classification task~\cite{wei2015hcp,chen2019multi, chen2019learning} for each snippet. For event-level, we follow~\cite{tian2020unified} and use F1-score as the evaluation metric. We do not use mAP as the event-level evaluation metric, as our framework is not an event proposal-based method. We set 4 sub-metric for each evaluation metric to evaluate the performance of different types of events and their average value, as shown in Tab.~\ref{table:mainexp}. For example, the audio sub-metric evaluates the localization performance of all events in the audio modality.

\textbf{Task Challenges.} Our proposed task takes several distinct challenges that should be addressed. First, a long form audio-visual video usually contains multiple events with various semantic categories, varies wildly in temporal length, and is depicted in huge different modalities. These issues practically play as barriers to achieving effective modeling of video content.
Second, modeling the inherent long-range dependencies is a key to understanding the full picture of a long video, but becoming more difficult when the scene in the long video changes dynamically or the audio modality and visual modality influence each other. Hence, a hit-to-the-point video understanding method needs to be proposed to step across these challenges.

\section{Method}
In this section, we propose an event-centric framework to solve the above challenges and aim to achieve a better understanding of long form audio-visual videos. Our framework contains three phases, \emph{i.e.,} snippet prediction in Sec.~\ref{ssec:5.1}, event extraction in Sec.~\ref{ssec:5.2}, and event interaction in Sec.~\ref{ssec:5.3}, as shown in Fig.~\ref{fig:framework}. 
We will give a concrete introduction in this section. \textcolor{black}{Due to the limitation of space, more details of the proposed framework are in the Sec. C of the \emph{Supp. Materials.}}

We use the pre-trained VGGish~\cite{gemmeke2017audio} model to extract audio features for each snippet, and use the pre-trained ResNet18~\cite{he2016deep} and R(2+1)D-18~\cite{tran2018closer} models to extract visual features for each snippet. The audio and visual featue are represented as $\{{a_t}\}^T_{t=1}$ and  $\{{v_t}\}^T_{t=1}$, respectively.

\subsection{Snippet Prediction}

\label{ssec:5.1}

\begin{wrapfigure}{r}{0.5\linewidth}

\centering
\vspace{-1em}
\includegraphics[width=1.0\linewidth]{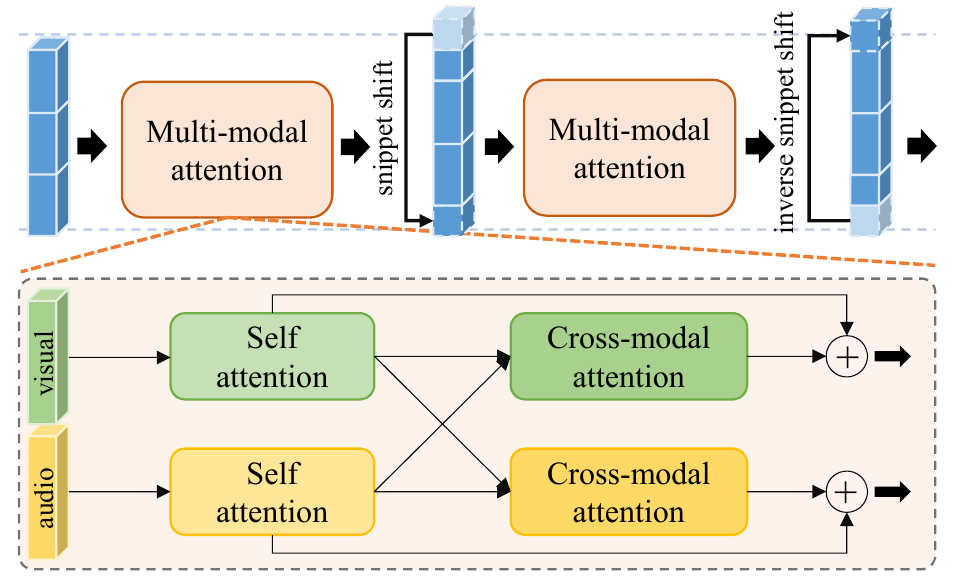}
\vspace{-1.5em}
\caption{\textbf{Architecture of the PMT layer.} \textbf{Top:} \textbf{The architecture of one PMT layer}, including two multimodal attention modules, a snippet shift operation, and an inverse snippet shift operation; \textbf{Bottom:} The detailed architecture of an attention module, including two self-attention units and two cross-modal attention units.}
\vspace{-1.5em}
\label{fig:phase1}

\end{wrapfigure}

In the proposed event-centric framework, we first focus on learning effective snippet features as the precondition of event localization. Concretely, we propose to capture the contained audio and visual events with different temporal lengths via pyramid window modules, and learn the interaction among the snippets of both modalities by multimodal attention module. These two modules constitute the \emph{Pyramid Multimodal Transformer} (PMT), as shown in Fig.~\ref{fig:phase1}.

\noindent
\textbf{Pyramid window module}. 
Inspired by the multiscale technique~\cite{he2015spatial,wang2021pyramid}, we use pyramid windows to capture events with different temporal scales then learn snippet features only within specific windows. Concretely, for the $l$-th PMT layer with a window size of $2^l$, the snippet features within the window are interacted with multimodal attention, then the outputs are used as the inputs of snippet interaction in the $(l+1)$-th layer, but with a window size of $2^{l+1}$, as the shown in Phase 1 of Fig.~\ref{fig:framework}.
In practice, we set six layers in the PMT, 
thus the window size ranges from 2 to 64.

\noindent
\textbf{Multimodal attention}.
To learn the interaction among snippets from the same or different modalities, we perform multimodal attention in each pyramid window. As shown in Fig.~\ref{fig:phase1}, each multimodal attention module contains four attention units, two of them are for audio and visual modeling, while the other two are for cross-modal interaction at the snippet level. Then, the updated snippet features are used as the inputs for the next layer.

To avoid the potential partition of an event caused by the pyramid window module\footnote
{Same-event snippets in different pyramid windows cannot fully interact with each other.
}, we also propose a kind of snippet shift strategy to capture the events across different pyramid windows. 
As shown in Fig.~\ref{fig:phase1}, we shift the top $2^l/2$ snippet features in the first window to the end of the video after the first multimodal attention module, then feed the updated snippet sequence to the second attention module. At the end of the PMT layer, we restore the temporal order of snippets and obtain the updated snippet features. With the stacked PMT layers, we could learn effective representation of audio and visual snippets, by considering the event with different temporal lengths and multimodal information. Then, we use a \emph{Temporal Attention Pooling} (TAP\footnote{More details about TAP are in \textcolor{black}{Sec. C.1 of} the \textit{Supp. Materials.}}) module to obtain the video-level and snippet-level prediction, its architecture is shown in the bottom left of Fig.~\ref{fig:framework}. The video-level prediction with binary cross-entropy loss is used for training this phase.

\subsection{Event Extraction}

\label{ssec:5.2}
For a long form audio-visual video, modeling its long-range dependency only at the snippet-level could not be effective because of its complex dynamic scenes and modality interaction. Hence, we propose to model the long-range dependency by excavating the unimodal events then exploring their inherent relations in long videos. 
In this section, we extract event features based on the preliminary snippet features and their category predictions from the snippet prediction phase, 
as Fig.~\ref{fig:framework} (\textit{Phase 2}).

\begin{table*}
\begin{center}
\vspace{-0.5em}
\caption{Experiment results of our framework and comparison methods. Ours (S) means our framework with only the snippet prediction phase. Ours (S+E) means our framework with the snippet prediction and event extraction phase. Ours (All) means our whole framework with all three phases. Avg. indicates the average value of the results in audio, visual, and audio-visual.}
\scalebox{0.88}{

\begin{tabular}{c|cccc|cccc}
    \toprule
	\multirow{2}*{Method} & \multicolumn{4}{c|}{F1-score (\%) (event-level )} & \multicolumn{4}{c}{mAP (\%) (snippet-level)} \\
	~ & Audio & Visual & Audio-Visual & Avg. & Audio & Visual & Audio-Visual & Avg.  \\
\midrule 

STPN~\cite{nguyen2018weakly} & 7.92 & 11.02 & 10.47 & 9.80 & 28.30 & 48.87 & 38.66 & 43.02\\

RSKP~\cite{huang2022weakly} & 7.64 & 8.39 & 11.79 & 9.27 & 29.42 & 47.43 & 22.25 & 33.03\\
 \hline
LongFormer~\cite{beltagy2020longformer} & 14.16 & 15.92 & 13.40 & 14.49 & 37.11 & 46.03 & 35.12 & 39.42\\

Transformer-LS~\cite{zhu2021long} & 16.55 & 22.41 & 17.87 & 18.94 & 37.91 & 50.10 & 37.77 & 41.92\\

ActionFormer~\cite{zhang2022actionformer} & 13.14 & 18.33 & 13.46 & 14.97 & 28.40 & 43.09 & 33.65 & 35.05\\

\hline
AVE~\cite{tian2018audio} & 16.42 & 11.97 & 13.71 & 13.71 & 42.41 & 47.44 & 39.84 & 43.23\\
AVSlowFast~\cite{xiao2020audiovisual} & 19.68 & 20.16 & 15.67 & 18.50 & 50.03 & 62.50 & 48.41 & 53.65 \\
HAN~\cite{tian2020unified} & 20.96 & 24.13 & 18.07 & 20.87 & 48.27 & 63.26 & 47.42 & 52.98\\
PSP~\cite{zhou2021positive} & 17.40 & 26.32 & 15.60 & 19.78 & 37.27 & 60.95 & 45.84 & 48.01\\
DHHN~\cite{jiang2022dhhn} & 21.44 & 18.82 & 14.23 & 18.16 & 49.95 & 59.59 & 47.74 & 52.42\\

\hline 
SlowFast~\cite{feichtenhofer2019slowfast} & 19.42 & 19.72 & 14.94 & 18.03 & 47.46 & 62.18 & 46.75 & 52.13 \\

MViT~\cite{fan2021multiscale} & 18.60 & 27.68 & 17.67 & 21.31 & 47.48 & \textbf{64.55} & 48.37 & 53.47\\
MeMViT~\cite{wu2022memvit} & 21.62 & 29.21 & 21.81 & 24.22 & 46.37 & 64.36 & 47.67 & 52.80\\

\hline \textbf{Ours (S)} & 24.81 & 30.26 & 23.06 & 26.04 & 48.76 & 62.26 & 47.00 & 52.67\\
\textbf{Ours (S+E)} & \textbf{25.76} & 31.46 & 25.36 & 27.53 & 49.31 & 63.38 & 47.97 & 53.55\\
\textbf{Ours (All)} & 25.36 & \textbf{32.44} & \textbf{26.61} & \textbf{28.14} & \textbf{51.66} & 64.10 & \textbf{50.11} & \textbf{55.29}\\ \bottomrule
\end{tabular}
}

\label{table:mainexp}

\end{center}
\vspace{-1.5em}
\end{table*}

An event usually consists of several snippets which are semantically related, but the remaining amounts of snippets could be in different semantics or backgrounds, especially for an extremely long snippet sequence. Hence, to obtain high-quality event representation as well as address the challenges in modeling the long-range dependency, we use an event-aware graph structure to model the long form video then refine event-aware snippet features. Concretely, we construct a graph for each event category of each modality, where each snippet is a node of the graph and the nodes are connected according to two kinds of edges: temporal edges and semantic edges. For temporal edges, we connect every two adjacent nodes, because successive snippets are usually considered to be similar in semantics. For semantic edges, the nodes are fully connected when they all have high confidence in the same event category. The confidence comes from the snippet-level predictions in the first phase. 

Based on the built event-aware graph, we propose to refine the event-aware snippet features via a graph attention based model.
Note that all the graphs share the same weights, either the kinds of modalities or the event categories they belong to. 
Then, based on the refined event-aware snippet features, we employ the TAP module to obtain their importance (\emph{i.e.,} attention weight) of the specific event, then perform weighted aggregation over the snippet features belonging to the same event to generate the final event feature.
Like the snippet prediction phase, we also use the binary cross-entropy loss to train this phase based on the prediction of the TAP module.

\subsection{Event Interaction}

\label{ssec:5.3}

As mentioned above, learning the relations among events is crucial in understanding long form videos, especially when faced with the requirements of modeling long-range dependency and different modalities. Considering these events are not independent with each other, we propose to learn the event relations by interacting events from two aspects, including intra-modal event interaction and cross-modal event interaction.
Based on the extracted event features in the second phase, event-aware self-attention and cross-modal attention at video-level are performed to explore the potential 
event relations, which are accordingly used to refine the event features. 
We then perform an event-level binary cross-entropy loss over the refined event features, under the video-level category label.

By cascading the introduced three phases, we could achieve an event-centric video understanding framework. \textcolor{black}{These three phases are jointly trained end-to-end,} and the third phase of event interaction is not considered during inference. \textcolor{black}{More details about the method are in Sec.C of the \emph{Supp. Materials}.}

\section{Experiments}

\subsection{Experimental Settings}

\label{ssec:6.1}

\noindent
\textbf{Implementation details.} We set batch size to 16 and use Adam to optimize the network, the initial learning rate of the three phases are 1e-4, 1e-4, and 2e-4, respectively. We train the model for 30 epochs, and the learning rate is reduced by a factor of 0.1 for every 10 epochs. The number of snippets of all videos is adjusted to 200 for effective training.

\noindent
\textbf{Comparison methods}. To validate the superiority of our proposed framework, we choose 13 related methods for comparison, including weakly supervised temporal action localization methods: STPN~\cite{nguyen2018weakly}, RSKP~\cite{huang2022weakly}; long sequence modeling methods: Longformer~\cite{beltagy2020longformer}, Transformer-LS~\cite{zhu2021long}, ActionFormer~\cite{zhang2022actionformer}; audio-visual learning methods: AVE~\cite{tian2018audio}, AVSlowFast~\cite{xiao2020audiovisual}, HAN~\cite{tian2020unified}, PSP~\cite{zhou2021positive}, DHHN~\cite{jiang2022dhhn}; video classification methods: SlowFast~\cite{feichtenhofer2019slowfast}, MViT~\cite{fan2021multiscale}, and MeMViT~\cite{wu2022memvit}.

\subsection{Results and Analysis}
\vspace{-0.5em}
\label{ssec:6.2}

\begin{table}[!t]
 \begin{minipage}{0.48\textwidth}
  \centering
    \makeatletter\def\@captype{table}\makeatother\caption{Ablation study on  feature interaction, where S-att represents self attention and C-att denotes cross-modal attention.}
    \scalebox{0.95}{
        \begin{tabular}{cc|cc}
        \hline
        \multicolumn{2}{c|}{Snippet interaction} & 
        \multicolumn{1}{c}{F1-score (\%) } & 
        \multicolumn{1}{c}{mAP (\%) } \\
        S-att & C-att & Avg. & Avg.  \\
        \midrule 
        \xmark & \xmark        &  9.66  & 38.71 \\
        $\checkmark$    & \xmark        &  21.45 & 48.63 \\
        \xmark          & $\checkmark$  &  17.45 & 51.85 \\
        $\checkmark$    & $\checkmark$  &  26.04 & 52.67 \\
        \hline
        \end{tabular}
        }
    \label{table:sniint}
    \vspace{-1.5em}
  \end{minipage}
  \begin{minipage}{0.04\textwidth}
      \centering
    \vspace{4cm} 
  \end{minipage}\hfill
  \begin{minipage}{0.48\textwidth}
   \centering
    \makeatletter\def\@captype{table}\makeatother\caption{Ablation study on event interaction, where S-att and C-att denote self and cross-modal attention, respectively.}
     \scalebox{0.95}{
        \begin{tabular}{cc|cc}
        \hline
        \multicolumn{2}{c|}{Event interaction} & 
        \multicolumn{1}{c}{F1-score (\%)} & 
        \multicolumn{1}{c}{mAP (\%)} \\
        S-att & C-att &  Avg. & Avg.  \\
        \midrule  
        \xmark & \xmark & 27.52 & 54.19\\
        \checkmark & \xmark & 27.65 & 54.79\\
        \xmark & \checkmark &  28.07  & 54.27\\
        \checkmark & \checkmark &  28.14  & 55.29\\
        \hline
        \end{tabular}
        }
    \label{table:eventint}   
    \vspace{-1.5em}
   \end{minipage}
\end{table}

\textbf{Comparison to other methods}. Experimental results of comparison methods and our methods are shown in Tab.~\ref{table:mainexp}. There are three points we could pay attention to.
Firstly, \textcolor{black}{temporal action localization~\cite{nguyen2018weakly,huang2022weakly} and long sequence modeling methods~\cite{beltagy2020longformer,zhu2021long,zhang2022actionformer} aim to effectively localize action events in untrimmed videos or model long sequences.} But they ignore the valuable cooperation among audio and video modality, which is important in achieving more comprehensive video event understanding. 
Secondly, although some methods~\cite{tian2018audio,zhou2021positive,tian2020unified,jiang2022dhhn} take the audio signal into account, they are consistently worse than our method. This could be because they mainly aim at understanding trimmed short videos, \textcolor{black}{resulting in limited modeling of long-range dependencies and event interactions.}
Thirdly, our proposed method outperforms all the comparison ones obviously, although some recent video classification methods~\cite{fan2021multiscale,wu2022memvit} achieve slightly better results on visual mAP, their overall performance still lags obviously behind our proposed method, showing that our proposed event-centric framework can localize both audio and visual events in long form audio-visual videos better. Additionally, we notice that localizing audio-visual events is more challenging because it requires precise localization in both modalities.

\noindent
\textbf{Effectiveness of three phases.} \textcolor{black}{As mentioned above,} our full method consists of three progressive phases. The performance of the snippet prediction phase has already surpassed most comparison methods, then the \textcolor{black}{subsequent phases} can further improve localization performance. Results are shown in the last three rows of Tab.~\ref{table:mainexp}, which indicate the potential importance of decoupling a long form audio-visual video into multiple uni-modal events with different lengths and modeling their inherent relations in both uni-modal and cross-modal scenarios.

\textbf{Effectiveness of feature interaction.} We further explore the effectiveness of snippet-level and event-level feature interaction. Results are shown in Tab.~\ref{table:sniint} and Tab.~\ref{table:eventint}, respectively.
We can get two conclusions from these results: \textbf{1)}
Both kinds of interactions benefit from intra-modal and cross-modal attention, which not only indicates the importance of the uni-modal sequence but also shows the meaning of cross-modal relations in achieving effective long form audio-visual video understanding. \textbf{2)} The single snippet prediction phase without any feature interaction can be viewed as a normal classification method, whose results are in the third row of Tab.~\ref{table:sniint}. Compared with our full model, its F1-score drop and mAP drop are 18.48\% and 16.58\%, respectively. This huge performance drop indicates that the inherent long-range dependency and cross-modal relations are crucial to solving our task. More experimental results and analysis are in \textcolor{black}{Sec. D.3 and D.4 of} the \emph{Supp. Materials}.

\section{Conclusion}
In this paper, we pose and tackle a challenging multisensory temporal event localization task in long form videos. We collect the LFAV dataset to facilitate our research and propose an event-centric framework to solve the task. Experimental results demonstrate that capturing multiple audio and visual events with different temporal lengths and modeling long-range dependencies by event interaction is crucial for the problem.
We hope our work could be a meaningful exploration towards more natural machine perception and bring some inspiration to audio-visual video understanding, \emph{e.g.}, audio-visual video dense captioning, reasoning over scene dynamics, \emph{etc}.

\medskip
{
\small
\bibliographystyle{ieee}
\bibliography{egbib}

}

\clearpage

\appendix

\section{Dataset Examples}
\textcolor{black}{Dataset examples can be found on the website of our project: \href{http://gewu-lab.github.io/LFAV/}{http://gewu-1ab.github.io/LFAV/}}

\section{Auxiliary Statistical Analysis}
\begin{figure*}[t]
\centering
\includegraphics[width=1\linewidth]{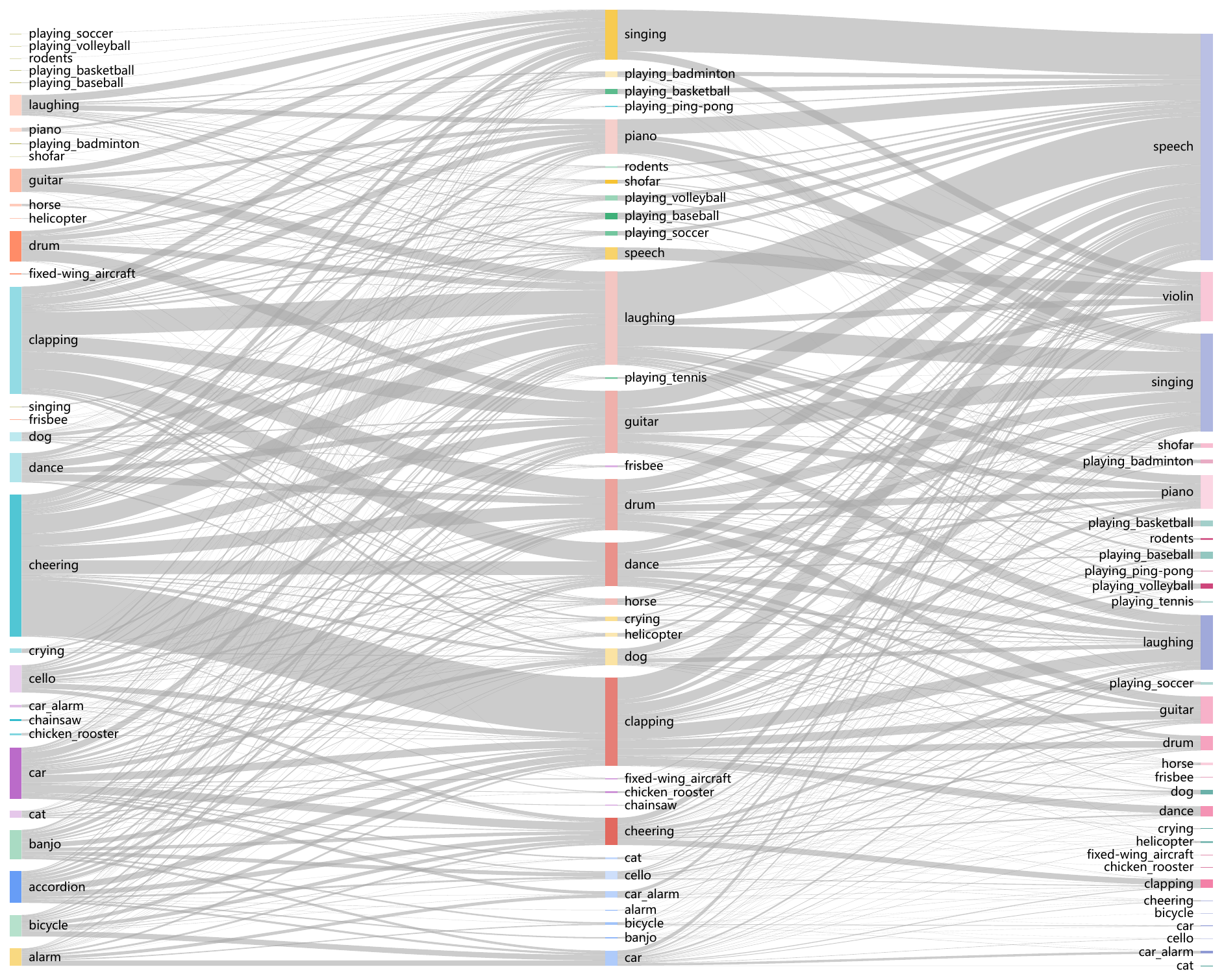}

\caption{Sankey diagram of video-level label in LFAV dataset, which shows the third-order interactions of video-level labels.}
\label{fig:sangji}

\end{figure*}

We show the number of video-level and event-level labels of each category in Tab.~\ref{table:video_level} and Tab.~\ref{table:event_level}, respectively. For video-level, the LFAV dataset contains 24,875 video-level annotations, including 11,404 visual event annotations and 13,471 audio event annotations, each category occuring in at least 146 videos. For event-level, the label distribution is similar to video-level (see Fig. 2(f) in the \emph{main paper}), but event-level labels just exist in the validation set and testing set. The LFAV dataset contains 23,666 event-level event annotations, including 11,331 visual event annotations and 12,335 audio event annotations. 
We also \textcolor{black}{show} the third-order interactions among different video-level label categories in Fig.~\ref{fig:sangji}. Almost all categories have dense interactions with other categories, and some of them have \textcolor{black}{closer} relations (\emph{e.g.}, \emph{clapping, laughing,} and \emph{speech}). These statistical results illustrate the diversity of the collected videos.

\begin{table*}[h]
\begin{center}

\caption{Number of video-level labels of each category. For each category, we stat number of videos that it occurs (\emph{i.e.}, occurs in at least one modality of the video), number of video-level visual labels, and number of video-level audio labels. For example, for category \emph{clapping}, it occurs in 1,486 videos in total, number of video-level visual labels and audio labels are 899 and 1,341, respectively.}
\vspace{0.5em}
\scalebox{0.7}{
\begin{tabular}{c|c|ccc|ccc|ccc|ccc}
\hline
& &   \multicolumn{3}{c|}{{\textbf{Training}}}    & 
    \multicolumn{3}{c|}{{\textbf{Validation}}}     & \multicolumn{3}{c|}{{\textbf{Testing}}}        
    & \multicolumn{3}{c}{{\textbf{Total}}}        \\ \cline{3-14} 
\multirow{-2}{*}{{\begin{tabular}[c]{@{}c@{}}\textbf{Label} \\ \textbf{id}\end{tabular}}} & 
\multirow{-2}{*}{{\textbf{Categories}}} & 
{\textbf{video}} & {\textbf{visual}} & {\textbf{audio}} & 
{\textbf{video}} & {\textbf{visual}} & {\textbf{audio}} & 
{\textbf{video}} & {\textbf{visual}} & {\textbf{audio}} & 
{\textbf{video}} & {\textbf{visual}} & {\textbf{audio}} \\ 
\hline
01 & speech     & 1972    & 626    & 1938      & 322    & 181    & 296    & 614    & 342    & 562    & 2908    & 1149    & 2796      \\
02 & clapping    & 951    & 564    & 851    & 158    & 95     & 148    & 377    & 240    & 342    & 1486    & 899    & 1341      \\
03 & cheering    & 1048    & 397    & 985    & 130    & 44     & 126    & 296    & 99     & 286    & 1474    & 540    & 1397      \\
04 & laughing    & 629    & 430    & 457    & 158    & 100    & 115    & 305    & 208    & 233    & 1092    & 738    & 805    \\
05 & singing    & 545    & 356    & 531    & 109    & 74     & 102    & 240    & 166    & 219    & 894    & 596    & 852    \\
06 & car        & 459    & 440    & 207    & 88     & 78     & 36    & 137    & 119    & 65    & 684    & 637    & 308    \\
07 & guitar     & 401    & 326    & 382    & 96     & 82     & 89    & 181    & 161    & 171    & 678    & 569    & 642    \\
08 & drum       & 377    & 181    & 349    & 65     & 40     & 61    & 132    & 80     & 124    & 574    & 301    & 534    \\
09 & piano      & 294    & 211    & 279    & 66     & 51     & 61    & 137    & 107    & 130    & 497    & 369    & 470    \\
10 & dance      & 255    & 240    & 81    & 44     & 43     & 3     & 113    & 108    & 24    & 412    & 391    & 108    \\
11 & alarm      & 347    & 255    & 331    & 28     & 15     & 24    & 33     & 14     & 28    & 408    & 284    & 383    \\
12 & dog        & 266    & 249    & 188    & 47     & 39     & 28    & 80     & 73     & 36    & 393    & 361    & 252    \\
13 & violin     & 234    & 192    & 225    & 48     & 46     & 47    & 82     & 78     & 80    & 364    & 316    & 352    \\
14 & playing basketball    & 267    & 267    & 179    & 13     & 14     & 11    & 32     & 32     & 26    & 312    & 313    & 216    \\
15 & playing badminton     & 193    & 192    & 124    & 26     & 26     & 25    & 50     & 49     & 45    & 269    & 267    & 194    \\
16 & horse      & 220    & 220    & 149    & 19     & 19     & 8     & 23     & 23     & 12    & 262    & 262    & 169    \\
17 & bicycle    & 194    & 191    & 24    & 21     & 19     & 3     & 47     & 41     & 17    & 262    & 251    & 44    \\
18 & cello      & 137    & 127    & 129    & 28     & 27     & 27    & 62     & 57     & 59    & 227    & 211    & 215    \\
19 & rodents    & 166    & 166    & 161    & 15     & 15     & 12    & 25     & 23     & 19    & 206    & 204    & 192    \\
20 & frisbee    & 184    & 181    & 90    & 6      & 5      & 2     & 12     & 11     & 3     & 202    & 197    & 95    \\
21 & fixed-wing aircraft   & 170    & 164    & 157    & 11     & 10     & 11    & 20     & 20     & 18    & 201    & 194    & 186    \\
22 & playing ping-pong     & 181    & 181    & 147    & 1      & 1      & 1     & 18     & 17     & 17    & 200    & 199    & 165    \\
23 & accordion     & 111    & 108    & 108    & 24     & 24     & 21    & 59     & 56     & 59    & 194    & 188    & 188    \\
24 & playing volleyball    & 147    & 147    & 71    & 14     & 14     & 9     & 28     & 27     & 17    & 189    & 188    & 97    \\
25 & playing baseball    & 151    & 151    & 50    & 7      & 7      & 6     & 31     & 30     & 12    & 189    & 188    & 68    \\
26 & cat        & 130    & 129    & 78    & 18     & 15     & 4     & 40     & 39     & 15    & 188    & 183    & 97    \\
27 & playing tennis     & 161    & 161    & 126    & 6      & 6      & 6     & 21     & 21     & 18    & 188    & 188    & 150    \\
28 & banjo      & 132    & 125    & 127    & 10     & 10     & 10    & 44     & 42     & 42    & 186    & 177    & 179    \\
29 & car\_alarm    & 149    & 107    & 140    & 12     & 8      & 8     & 18     & 6      & 14    & 179    & 121    & 162    \\
30 & helicopter    & 100    & 99     & 74    & 28     & 28     & 27    & 50     & 49     & 47    & 178    & 176    & 148    \\
31 & crying       & 136    & 121    & 117    & 13     & 8      & 10    & 21     & 18     & 16    & 170    & 147    & 143    \\
32 & chainsaw    & 105    & 104    & 100    & 26     & 24     & 26    & 35     & 34     & 35    & 166    & 162    & 161    \\
33 & playing soccer     & 118    & 117    & 61    & 17     & 17     & 13    & 26     & 26     & 13    & 161    & 160    & 87    \\
34 & chicken\_rooster     & 135    & 122    & 131    & 4      & 4      & 3     & 11     & 10     & 8     & 150    & 136    & 142    \\
35 & shofar     & 118    & 116    & 107    & 15     & 14     & 14    & 13     & 12     & 12    & 146    & 142    & 133    \\ \hline
 & \textbf{Total}   & \textbf{11183}    & \textbf{7763}     & \textbf{9254}     & 
                      \textbf{1693}     & \textbf{1203}     & \textbf{1393}     & 
                      \textbf{3413}     & \textbf{2438}     & \textbf{2824}     & 
                      \textbf{16289}    & \textbf{11404}    & \textbf{13471}        
\\ \hline
\end{tabular}

}

\label{table:video_level}
\end{center}

\end{table*}
\clearpage

\begin{table}[t]
\begin{center}
\caption{Number of event-level labels of each category. And the label id is the same as in Tab.~\ref{table:video_level}.}
\vspace{0.5em}
\scalebox{0.9}{

\begin{tabular}{c|cc|cc|cc}
\hline
{\color[HTML]{000000} }      & \multicolumn{2}{c|}{{\color[HTML]{333333} \textbf{Validation}}}     & \multicolumn{2}{c|}{{\color[HTML]{333333} \textbf{Testing}}}           & \multicolumn{2}{c}{{\color[HTML]{333333} \textbf{Total}}}           \\ \cline{2-7} 
\multirow{-2}{*}{{\color[HTML]{000000} \textbf{Categories}}} & {\color[HTML]{333333} \textbf{visual}} & {\color[HTML]{333333} \textbf{audio}} & {\color[HTML]{333333} \textbf{visual}} & {\color[HTML]{333333} \textbf{audio}} & {\color[HTML]{333333} \textbf{visual}} & {\color[HTML]{333333} \textbf{audio}} \\ \hline
speech    & 629    & 1151    & 1255     & 2217    & 1884     & 3368    \\
clapping          & 275    & 437     & 662    & 1116    & 937    & 1553    \\
cheering          & 132    & 469     & 280    & 1024    & 412    & 1493    \\
laughing          & 338    & 394     & 597    & 831     & 935    & 1225    \\
singing           & 237    & 259     & 655    & 556     & 892    & 815     \\
car       & 269    & 55    & 373    & 163     & 642    & 218     \\
guitar    & 274    & 137     & 670    & 356     & 944    & 493     \\
drum      & 156    & 92    & 372    & 234     & 528    & 326     \\
piano     & 131    & 123     & 339    & 233     & 470    & 356     \\
dance     & 105    & 4     & 225    & 32    & 330    & 36    \\
alarm     & 28     & 66    & 41     & 63    & 69     & 129     \\
dog       & 100    & 64    & 207    & 137     & 307    & 201     \\
violin    & 232    & 95    & 239    & 178     & 471    & 273     \\
playing basketball           & 25     & 13    & 101    & 68    & 126    & 81    \\
playing badminton            & 52     & 68    & 109    & 168     & 161    & 236     \\
horse     & 44     & 12    & 50     & 17    & 94     & 29    \\
bicycle           & 70     & 3     & 110    & 43    & 180    & 46    \\
cello     & 145    & 46    & 188    & 95    & 333    & 141     \\
rodents           & 27     & 27    & 25     & 26    & 52     & 53    \\
frisbee           & 9      & 2     & 38     & 7     & 47     & 9     \\
fixed-wing aircraft          & 32     & 27    & 36     & 23    & 68     & 50    \\
playing ping-pong            & 11     & 11    & 34     & 91    & 45     & 102     \\
accordion         & 67     & 33    & 147    & 93    & 214    & 126     \\
playing volleyball           & 72     & 70    & 154    & 118     & 226    & 188     \\
playing baseball             & 8      & 7     & 77     & 23    & 85     & 30    \\
cat       & 50     & 16    & 75     & 32    & 125    & 48    \\
tennis    & 9      & 6     & 53     & 38    & 62     & 44    \\
banjo     & 38     & 14    & 117    & 135     & 155    & 149     \\
car alarm         & 16     & 17    & 23     & 36    & 39     & 53    \\
helicopter        & 46     & 46    & 85     & 66    & 131    & 112     \\
crying      & 15     & 14    & 28     & 29    & 43     & 43    \\
chainsaw          & 64     & 48    & 62     & 69    & 126    & 117     \\
playing soccer    & 38     & 20    & 88     & 66    & 126    & 86    \\
chicken\_rooster             & 4      & 14    & 21     & 31    & 25     & 45    \\
shofar    & 18     & 19    & 29     & 42    & 47     & 61    \\ \hline
\textbf{Total}    & \textbf{3766}       & \textbf{3879}      & \textbf{7565}       & \textbf{8456}      & \textbf{11331}      & \textbf{12335}     \\ \hline
\end{tabular}

}

\label{table:event_level}
\end{center}

\end{table}
\clearpage

\clearpage

\section{Auxiliary Explanation of the Method}
\subsection{Temporal Attention Pooling}
We can obtain the video-level and snippet-level prediction by the \emph{Temporal Attention Pooling} (TAP) module. Inputs of the TAP module are audio features $\{{a_t}\}^T_{t=1}$ and visual features $\{{v_t}\}^T_{t=1}$, then outputs of the TAP module are video-level audio prediction ${p^a}$, video-level visual prediction ${p^v}$, snippet-level audio prediction $\{{p^a_t}\}^T_{t=1}$, and snippet-level visual prediction $\{{p^v_t}\}^T_{t=1}$. Video-level predictions are used to train the model, snippet-level predictions are used to construct event graphs during training and evaluate model performance during validation and testing. Snippet-level predictions \textcolor{black}{are} calculated as:
\begin{equation}
\setlength{\arraycolsep}{1.0pt}
    p^a_t=Sigmoid(FC(a_t)),
\end{equation}
\begin{equation}
\setlength{\arraycolsep}{1.0pt}
    p^v_t=Sigmoid(FC(v_t)),
\end{equation}

where $FC$ represents a fully connected layer, the audio modality and visual modality share the same fully connected layer. For video-level predictions, another fully connected layer $FC^{'}$ is used to obtain normalized attention weights of each audio and visual snippet at first:
\begin{equation}
\setlength{\arraycolsep}{1.0pt}
    w^a_t=\frac{exp(FC^{'}(a_t))}{\sum_{j=1}^Texp(FC^{'}(a_j))},
\end{equation}
\begin{equation}
\setlength{\arraycolsep}{1.0pt}
    w^v_t=\frac{exp(FC^{'}(v_t))}{\sum_{j=1}^Texp(FC^{'}(v_j))},
\end{equation}
where $w^a_t$ represents the weight of $t$-th audio snippet and $w^v_t$ represents the weight of $t$-th visual snippet. Two modalities also share the same fully connected layer. Then video-level predictions 
\textcolor{black}{are} calculated as:
\begin{equation}
\setlength{\arraycolsep}{1.0pt}
    p^a=\sum_{t=1}^Tw^a_t\odot p^a_t,
\end{equation}
\begin{equation}
\setlength{\arraycolsep}{1.0pt}
    p^v=\sum_{t=1}^Tw^v_t\odot p^v_t,
\end{equation}
where $\odot$ represents the element-wise multiplication.

\begin{figure}[t]

\centering
\includegraphics[width=0.7\linewidth]{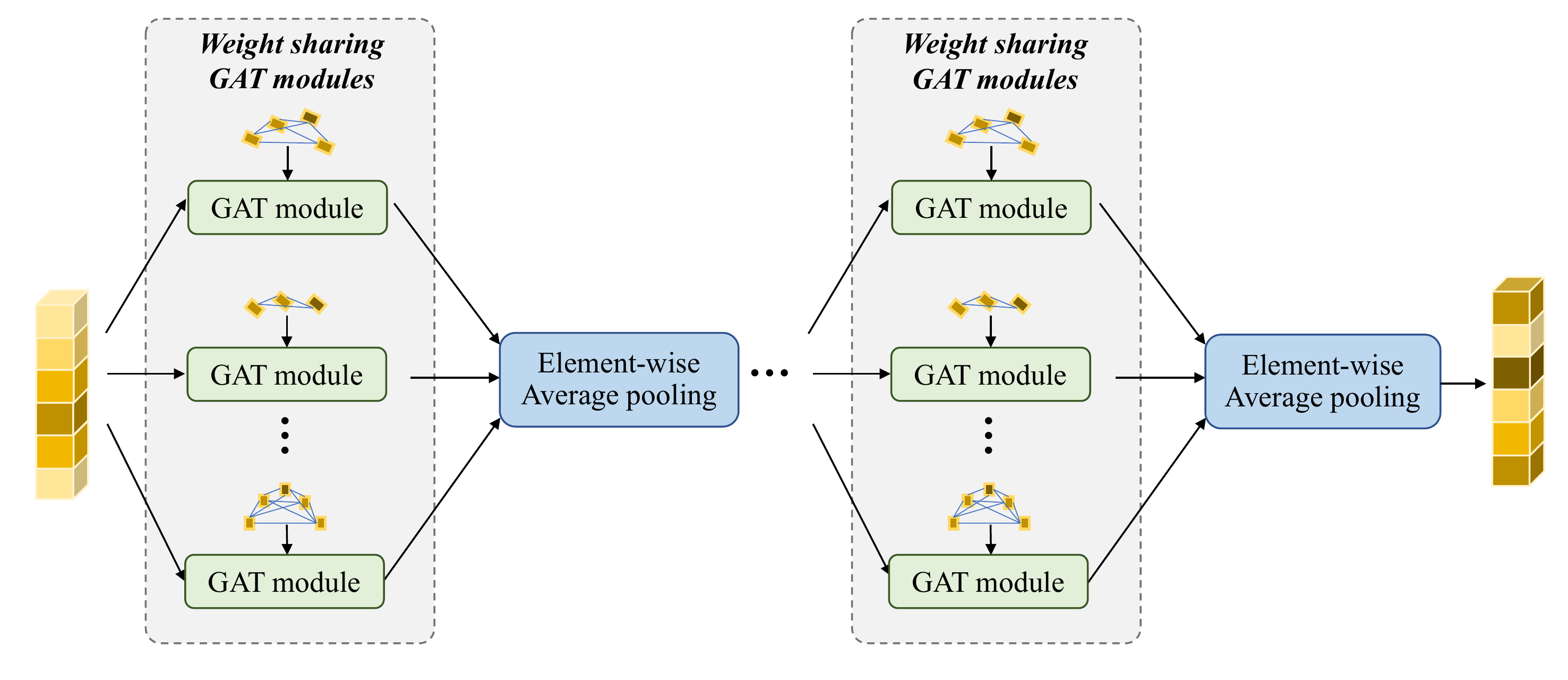}
\caption{\textbf{The architecture of the GAT based model.} \textcolor{black}{Model architecture of audio and visual modalities are the same, this figure shows the model architecture of one modality.}}
\label{fig:graph}
\vspace{0.5em}
\end{figure}

Compared with the MMIL Pooling in HAN~\cite{tian2020unified}, we do not perform modality-wise attention because we have independent audio and visual labels during training.

\subsection{Graph Attention Network Based Model}
We propose a graph attention network (GAT)~\cite{velivckovic2018graph} based model to refine event-aware snippet features, Fig.~\ref{fig:graph} shows its detailed architecture. For a layer in the model, all GAT modules in the same layer share the same weights but use category-aware graph structures to aggregate snippet features in different events. \textcolor{black}{For each GAT module, we obtain refined event-aware snippet features. Then the output of the layer is the average output of all GAT modules.}

\subsection{Snippet Reweighting}
In the event interaction phase, we use refined event features to reweight each snippet, then obtain updated video-level predictions for training. Concretely, the attention weight of each snippet \textcolor{black}{is} reset according to the cosine similarity between the event feature and the snippet feature. Suppose $\hat{a}^i$ and $\hat{v}^i$ are event features of the $i$-th category of the audio modality and the visual modality, respectively. For the $i$-th category of the audio modality, weight of the $t$-th snippet is the $i$-th element of $w^a_t$, it \textcolor{black}{is} represented as $w^a_t[i]$ and recalculated as: 
\begin{equation}
\setlength{\arraycolsep}{1.0pt}
    w^a_t[i]=\frac{exp(cos(\hat{a}^i,a_t))}{\sum_{j=1}^Texp(cos(\hat{a}^i,a_j))},
\end{equation}
where $cos$ represents the cosine similarity function, then the updated video-level prediction of $i$-th category of the audio modality \textcolor{black}{is} calculated as:
\begin{equation}
\setlength{\arraycolsep}{1.0pt}
    p^a[i]=\sum_{t=1}^Tp^a_t[i]\odot w^a_t[i].
\end{equation}
For the visual modality, the same calculation will be also performed to update video-level prediction. For categories that have not extracted events, we do not reweight snippets and keep previous video-level predictions.

\subsection{Loss Function}
Suppose ${p^a_{(j)}}$ and ${p^v_{(j)}}$ represent video-level audio and video predictions of the $j$-th phase, respectively. For each phase, a binary cross-entropy loss is performed to optimize the model:
\begin{equation}
\setlength{\arraycolsep}{1.0pt}
    \mathcal{L}_j=BCE({p^a_{(j)}},y^a)+BCE({p^v_{(j)}},y^v), ~j=1,2,3,
\end{equation}
where $y^a$ and $y^v$ are audio and visual event labels, respectively. $\mathcal{L}_j$ is the video-level loss of the $j$-th phase. An event-level binary cross-entropy loss (event loss) is also performed in the event interaction phase. Event predictions \textcolor{black}{are} first calculated as:
\begin{equation}
\setlength{\arraycolsep}{1.0pt}
    \hat{p}^{a,i}=Sigmoid(FC(\hat{a}^i))[i],
\end{equation}
\begin{equation}
\setlength{\arraycolsep}{1.0pt}
    \hat{p}^{v,i}=Sigmoid(FC(\hat{v}^i))[i],
\end{equation}
where $\hat{p}^{a,i}$ and $\hat{p}^{v,i}$ are audio event prediction and visual event prediction of the $i$-th category, $FC$ is the fully connected layer which used to obtain snippet-level predictions in the TAP module of event extraction phase. Suppose $n_a$ is the number of extracted audio events, $n_v$ is the number of extracted visual events, $S_a$ and $S_v$ are sets of category indexes of extracted events in two modalities, respectively. Then the event loss $\mathcal{L}_e$ \textcolor{black}{is} calculated as:  
\begin{equation}
\setlength{\arraycolsep}{1.0pt}
    \mathcal{L}_{ea}=\frac{1}{n_a}\sum_{i\in S_a}BCE(\hat{p}^{a,i},y^a[i]),
\end{equation}
\begin{equation}
\setlength{\arraycolsep}{1.0pt}
    \mathcal{L}_{ev}=\frac{1}{n_v}\sum_{i\in S_v}BCE(\hat{p}^{v,i},y^v[i]),
\end{equation}
\begin{equation}
\setlength{\arraycolsep}{1.0pt}
    \mathcal{L}_e=\mathcal{L}_{ea}+\mathcal{L}_{ev},
\end{equation}
the total training loss is:
\begin{equation}
\setlength{\arraycolsep}{1.0pt}
    \mathcal{L}=\mathcal{L}_1+\mathcal{L}_2+\mathcal{L}_3+\alpha \mathcal{L}_e,
\end{equation}
where $\alpha$ is the weight of event loss.

\section{Auxiliary Experimental Results}
\subsection{Detailed Training Settings}

The detailed training settings are shown in Tab.~\ref{table:trainset}.

\subsection{Settings of Comparison Methods}
\begin{wraptable}{rb}{7.7cm}
\begin{center}
\vspace{-1em}
\caption{Detailed training settings of the framework, the learning rate step is set to (10; 0.1), means the learning rate is reduced by a factor of 0.1 for every 10 epochs.}

\label{table:trainset}
\scalebox{0.87}{
\begin{tabular}{c|c}
\toprule
Hyperparameter & Value\\
\midrule 
batch size & 16 \\
train epochs & 30 \\
initial learning rate (snippet prediction phase) & 1e-4 \\
initial learning rate (event extraction phase) & 1e-4 \\
initial learning rate (event interaction phase) & 2e-4 \\
learning rate step & 10; 0.1 \\
event loss weight & 0.3 \\ 
feature dim & 512 \\ 
depth of PMT & 6 \\ 
number of heads of PMT & 4 \\
dropout ratio of PMT & 0.2 \\
confidence threshold of semantic edges & 0.5 \\
depth of graph network & 2 \\
number of heads of graph network & 1 \\
dropout ratio of graph network & 0 \\
dropout ratio of event interaction layers & 0 \\
\bottomrule
\end{tabular}
}
\end{center}
\vspace{-1em}
\end{wraptable}

We modify some of the comparison methods to make them possible to solve the multisensory temporal event localization task.  For STPN~\cite{nguyen2018weakly}, the weight of sparse loss is set to 0 to achieve better performance. For Longformer~\cite{beltagy2020longformer} and
Transformer-LS~\cite{zhu2021long}, we use HAN as the backbone and replace attention layers in the HAN model with attention layers in Longformer or Transformer-LS, their window size is set to 50. For ActionFormer~\cite{zhang2022actionformer}, we use its backbone and FPN decoder to extract multi-scale features and use the upsampling strategy proposed in U-Net~\cite{ronneberger2015u} to fuse multi-scale features. The number of transformer blocks is set to 5 and the last 3 transformer layers are downsampling layers. For Longformer, Transformer-LS, and ActionFormer, the number of snippets of all videos is adjusted to 200 while testing. For methods proposed for audio-visual event localization~\cite{tian2018audio} or audio-visual video parsing ~\cite{tian2020unified} tasks  AVE~\cite{tian2018audio}, 
PSP~\cite{zhou2021positive}, 
HAN~\cite{tian2020unified}, and DHHN~\cite{jiang2022dhhn}, we use both audio and visual labels for training and use the same loss to HAN.  For all comparison methods, predictions of audio-visual events are the multiplication of \textcolor{black}{audio events predictions and visual events predictions}~\cite{tian2020unified}.

\subsection{\textcolor{black}{More Ablation Studies}}

\begin{wrapfigure}{r}{0.6\linewidth}
\centering
\vspace{-1em}
\includegraphics[width=1.0\linewidth]{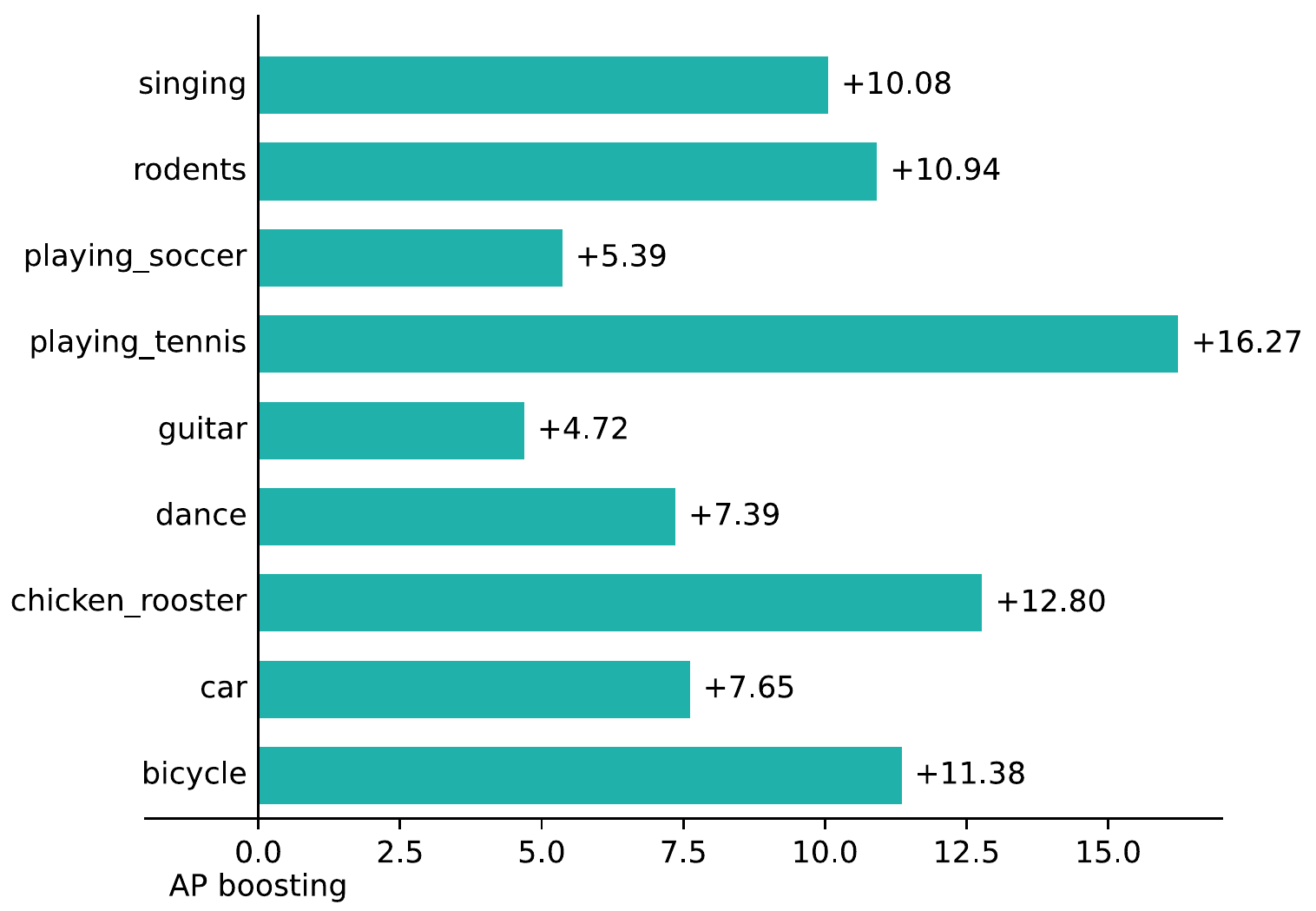}

\caption{AP boosting of some typical categories when considering the joint learning of audio and visual.}

\vspace{-0.5em}
\label{fig:ap_class}
\end{wrapfigure}

\textbf{Effectiveness of Audio-visual Joint Training.} As mentioned in the \emph{main paper}, ablation studies about feature interaction have shown that cross-modal attention improves video understanding through audio and visual cooperation. To further illustrate it, \textcolor{black}{we show the AP boosting of typical categories when considering snippet-level cross-modal attention in} Fig.~\ref{fig:ap_class}. Audio and visual take distinct viewpoints in describing the video content but facilitate the understanding of the video cooperatively. For example, for category \emph{singing}, audio provides the rhythm of the song, and visual provides the pose of the singer; for category \emph{playing\_tennis}, audio provides the impact sound of tennis, and visual provides the motion of players.  

\begin{wrapfigure}{r}{0.6\linewidth}
\centering
\vspace{-1em}
\includegraphics[width=1.0\linewidth]{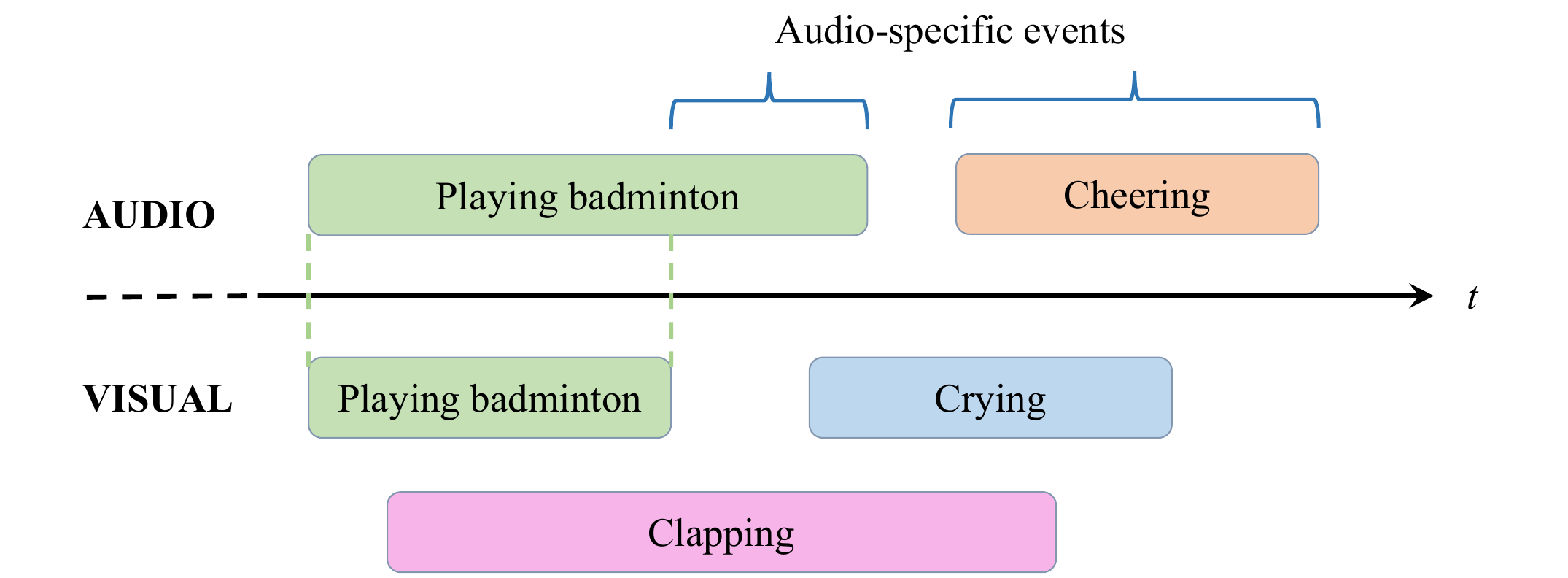}
\caption{An example of audio-specific events.}
\label{fig:audio_specific}
\vspace{-1.0em}
\end{wrapfigure}

\textbf{\textcolor{black}{Comparison of Audio Supervision.}} \textcolor{black}{We use visual annotations as audio supervision to train our model (\emph{i.e.}, regard visual labels as both audio and visual supervision during training). The performance of localizing audio-specific events (\emph{i.e.}, events only occur in the audio modality and do not occur in the visual modality, as shown in Fig.~\ref{fig:audio_specific}) with different audio supervision are in Tab.~\ref{table:annotation}. 
When we view visual annotations as audio supervision, the snippet-level recall of audio-specific events is even far below 50\%, showing that mismatched audio supervision will result in the collapse of audio-specific events localization performance.}
\begin{wraptable}{r}{7.0cm}
\vspace{-2.2em}
\begin{center}
\caption{Ablation study on audio supervision. We choose snippet-level recall as the metric to achieve a more accurate localization performance of audio-specific events, the threshold is set to 0.5.} 
\begin{tabular}{c|cc}
    \toprule
Audio supervision &  Recall (\%) \\
\midrule Visual annotations & 30.60 \\Audio annotations & 63.09 \\
\bottomrule 
\end{tabular}

\label{table:annotation}

\vspace{-1.5em}

\end{center}

\end{wraptable}

\textbf{\textcolor{black}{Effectiveness of Snippet Shift.}} We explore the effectiveness of the snippet shift strategy in the snippet prediction phase, results are shown in Tab.~\ref{table:absshift}. The snippet shift strategy is beneficial to event-aware snippet interaction. Notice that snippet shift is a computation-efficient operation, hence, the performance improvement from snippet shift is almost computationally free.

\textbf{\textcolor{black}{Effectiveness of Event Loss.}} \textcolor{black}{We also explore the effectiveness of event loss in the event interaction phase. The event loss is performed over snippets of extracted events. Corresponding results are shown in Tab.~\ref{table:event_loss}, which indicates the importance of event loss in localizing multiple events in long form videos.}

\subsection{Visualization Results}

\begin{figure*}[!t]
\centering
\includegraphics[width=1.0\linewidth]{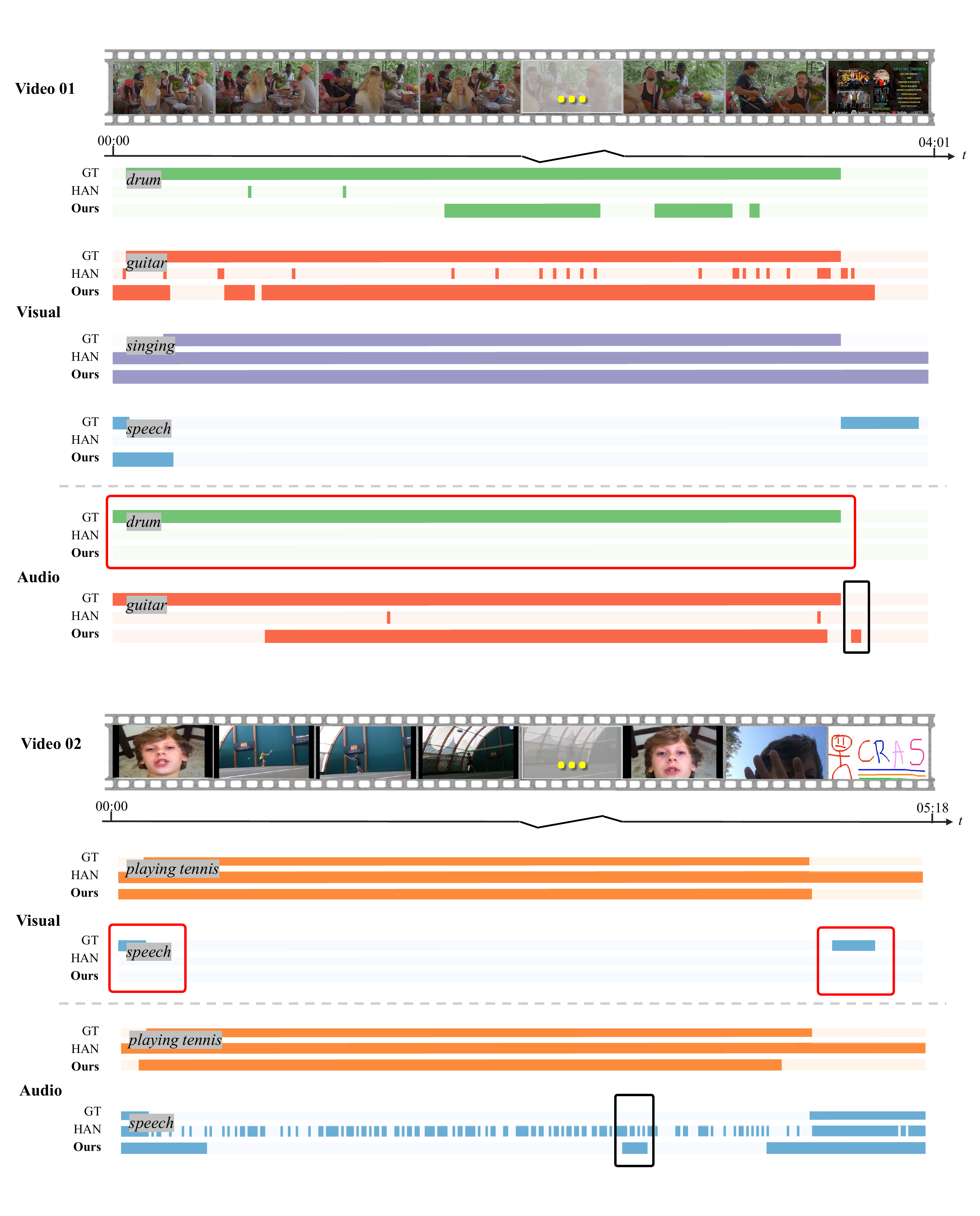}
\vspace{-0.5em}
\caption{Two examples of visualized localization results of our method and the best comparison method HAN, red rectangles in the figure show some failure cases of our method.}
\label{fig:vis}
\vspace{-5mm}
\end{figure*}

We visualize the event-level localization results in the videos, two examples are shown in Fig.~\ref{fig:vis}. Compared with the audio-visual video parsing method HAN~\cite{tian2020unified}, our proposed method achieves better localization results. In some situations (\emph{e.g.}, event \emph{guitar} in both audio and visual modality of \emph{video 01}, and event \emph{speech} in the audio modality of \emph{video 02}), HAN tends to localize some sparse and short video clips instead of a long and complete event, which shows that HAN exists some limitations to understanding long-form videos. The possible reason is that HAN cannot learn long-range dependencies well. 

We also notice that, although our proposed event-centric method has achieved the best performance among all methods, there still exist some failure cases in the shown examples. The auditory event of \emph{drum} in \emph{video 01} and visual event of \emph{speech} in \emph{video 02} are not well localized by both methods, including ours (marked with red box). There are also existing some short and sparse clips in our prediction results (marked with the black box). We can find that these multisensory events take huge different lengths and occur in a dynamic long-range scene, which makes multisensory temporal event localization become a very challenging task, especially with only video-level labels in training. Although our method has partially addressed it according to the shown results, this challenging task still needs more exploration in future work (\emph{e.g.}, considering some post-processing methods for the prediction results to merge or correct short and sparse prediction clips).

\clearpage

\begin{table}[!t]
 \begin{minipage}{0.5\textwidth}
  \centering
    \makeatletter\def\@captype{table}\makeatother\caption{Ablation study on snippet shift strategy in the snippet prediction phase.}

            \begin{tabular}{c|cc}
        \toprule
        \multirow{2}*{Snippet shift}  & 
        \multicolumn{1}{c}{F1-score}  & 
        \multicolumn{1}{c}{mAP} \\
        	~  &  Avg. & Avg.  \\
        \midrule 
        \xmark &  25.98  & 52.51\\
        \checkmark & 26.04  & 52.67\\
        \bottomrule 
        \end{tabular}
    \label{table:absshift}
    \vspace{-0.5em}
  \end{minipage}
  \begin{minipage}{0.5\textwidth}
   \centering
        \makeatletter\def\@captype{table}\makeatother\caption{Ablation study on event loss.}
     
          \begin{tabular}{c|cc}
            \toprule
            \multirow{2}*{Event loss}  & 
            \multicolumn{1}{c}{F1-score}  & 
            \multicolumn{1}{c}{mAP} \\
            	~  &  Avg. & Avg.  \\
            \midrule 
            \xmark &  27.64  & 54.70\\
            \checkmark & 28.14  & 55.29\\
            \bottomrule 

      \end{tabular}
    \label{table:event_loss}   
    \vspace{-0.5em}
   \end{minipage}
\end{table}

\clearpage

\section{\textcolor{black}{Discussion}}

As mentioned in the \emph{main paper}, we hope our work could bring some inspiration to further research of audio-visual video understanding, here we give a brief discussion. 

\noindent
\textbf{Audio-visual video dense captioning.}
The goal of the video dense captioning task is to detect multiple events in videos then describe them using natural language sentences~\cite{krishna2017dense}. The audio was exploited in previous video dense captioning approaches~\cite{iashin2020better,iashin2020multi,rahman2019watch}, while all of them did not treat audio as an independent modality. They used visual-level annotations to label the audio modality. However, we want to note that there are diverse audio and visual events and the two modalities are not always temporally correlated. Accurate multisensory temporal event localization results can help us solve a "real" audio-visual video dense captioning problem that aims to detect and describe different audio and visual events in long form videos.

\noindent
\textbf{Reasoning over scene dynamics.} Causal relations between events are ubiquitous in the real world (\emph{e.g.}, In \emph{Fig. 1 in the \emph{main paper}}, the badminton player cries because he has won the game and he is very excited.). Due to the multisensory characteristic of events, data in different sensory channels also exist in causal relations. These casual relations can be a bridge to achieving a high-level understanding of the video. Humans usually can reason about causal relations between events easily, but for the machine, \emph{understanding} and \emph{reasoning} are essential difficult tasks. Our proposed event-centric framework can capture scene dynamics in videos, which serves as the cornerstone to facilitate future research in complex multisensory scene reasoning.

\end{document}